\input harvmac.tex

\input epsf.tex
\def\figin{\epsfcheck\figin}\def\figins{\epsfcheck\figins}
\def\epsfcheck{\ifx\epsfbox\UnDeFiNeD
\message{(NO epsf.tex, FIGURES WILL BE IGNORED)}
\gdef\figin##1{\vskip2in}\gdef\figins##1{\hskip.5in}
\else\message{(FIGURES WILL BE INCLUDED)}%
\gdef\figin##1{##1}\gdef\figins##1{##1}\fi}
\def\DefWarn#1{}
\def\figinsert{\goodbreak\midinsert}
\def\ifig#1#2#3{\DefWarn#1\xdef#1{fig.~\the\figno}
\writedef{#1\leftbracket fig.\noexpand~\the\figno}%
\figinsert\figin{\centerline{#3}}\medskip\centerline{\vbox{\baselineskip12pt
\advance\hsize by -1truein\noindent\footnotefont{\bf Fig.~\the\figno:} #2}}
\bigskip\endinsert\global\advance\figno by1}


\def\Z{{\bf Z}}

\def\K3{{\bf K3}}
\def\journal#1&#2(#3){\unskip, \sl #1\ \bf #2 \rm(19#3) }
\def\andjournal#1&#2(#3){\sl #1~\bf #2 \rm (19#3) }

\def\bar{\overline}
\def\hat{\widehat}

\def\tilde{\widetilde}

\def\frac#1#2{{#1\over#2}}

\def\ket#1{|#1\rangle}
\def\bra#1{\langle#1|}
\def\vev#1{\langle#1\rangle}

\def\inbar{\,\vrule height1.5ex width.4pt depth0pt}
\def\IC{\relax\hbox{$\inbar\kern-.3em{\rm C}$}}
\def\IR{\relax{\rm I\kern-.18em R}}
\def\IP{\relax{\rm I\kern-.18em P}}
\def\Z{{\bf Z}}

%
%


%
\catcode`\@=11
\def\slash#1{\mathord{\mathpalette\c@ncel{#1}}}
\overfullrule=0pt

\def\CC{{\cal C}}

\def\HH{{\cal H}}

\def\NN{{\cal N}}

\def\lam{\lambda}

\def\underrel#1\over#2{\mathrel{\mathop{\kern\z@#1}\limits_{#2}}}

\catcode`\@=12


%
\def\bra#1{\left\langle #1\right|}
\def\ket#1{\left| #1\right\rangle}
\def\vev#1{\left\langle #1 \right\rangle}
\def\det{{\rm det}}

\def \sinh{{\rm sinh}}
\def \cosh{{\rm cosh}}

\def\det{{\rm det}}
\def\exp{{\rm exp}}


\def \ov {\over}
\def \p {\partial}
\def \ha {{1 \ov 2}}
\def \al {\alpha}
\def \lam {\lambda}
\def \sig {\sigma}

\def \ep {\epsilon}

\def \apr {\alpha'}

\def\IL{\relax{\rm I\kern-.18em L}}
\def\IH{\relax{\rm I\kern-.18em H}}
\def\IR{\relax{\rm I\kern-.18em R}}
\def\IC{\relax\hbox{$\inbar\kern-.3em{\rm C}$}}





\def\makeblankbox#1#2{\hbox{\lower\dp0\vbox{\hidehrule{#1}{#2}%
   \kern -#1
   \hbox to \wd0{\hidevrule{#1}{#2}%
      \raise\ht0\vbox to #1{}
      \lower\dp0\vtop to #1{}
      \hfil\hidevrule{#2}{#1}}%
   \kern-#1\hidehrule{#2}{#1}}}%
}%
\def\hidehrule#1#2{\kern-#1\hrule height#1 depth#2 \kern-#2}%
\def\hidevrule#1#2{\kern-#1{\dimen0=#1\advance\dimen0 by #2\vrule
    width\dimen0}\kern-#2}%
\def\openbox{\ht0=1.2mm \dp0=1.2mm \wd0=2.4mm  \raise 2.75pt
\makeblankbox {.25pt} {.25pt}  }

\def\bun#1/#2{\leavevmode
   \kern.1em \raise .5ex \hbox{\the\scriptfont0 #1}%
   \kern-.1em $/$%
   \kern-.15em \lower .25ex \hbox{\the\scriptfont0 #2}%
}

\def\opensquare{\ht0=3.4mm \dp0=3.4mm \wd0=6.8mm  \raise 2.7pt
\makeblankbox {.25pt} {.25pt}  }


\def\sector#1#2{\ {\scriptstyle #1}\hskip 1mm
\mathop{\opensquare}\limits_{\lower
1mm\hbox{$\scriptstyle#2$}}\hskip 1mm}

\def\tsector#1#2{\ {\scriptstyle #1}\hskip 1mm
\mathop{\opensquare}\limits_{\lower
1mm\hbox{$\scriptstyle#2$}}^\sim\hskip 1mm}




\lref\GutperleAI{
M.~Gutperle and A.~Strominger,
``Spacelike branes,''
JHEP {\bf 0204}, 018 (2002)
[arXiv:hep-th/0202210].
}

\lref\ZwiebachIE{
B.~Zwiebach,
``Closed string field theory: Quantum action and the B-V master equation,''
Nucl.\ Phys.\ B {\bf 390}, 33 (1993)
[arXiv:hep-th/9206084].
}

\lref\DiVecchiaPR{
P.~Di Vecchia, M.~Frau, I.~Pesando, S.~Sciuto, A.~Lerda and R.~Russo,
``Classical p-branes from boundary state,''
Nucl.\ Phys.\ B {\bf 507}, 259 (1997)
[arXiv:hep-th/9707068].
}

\lref\zuber{
C.~Itzykson and J.~B.~Zuber,
``Quantum Field Theory,''
}

\lref\GarousiTR{
M.~R.~Garousi,
``Tachyon couplings on non-BPS D-branes and Dirac-Born-Infeld action,''
Nucl.\ Phys.\ B {\bf 584}, 284 (2000)
[arXiv:hep-th/0003122].
}

\lref\SenMD{
A.~Sen,
``Supersymmetric world-volume action for non-BPS D-branes,''
JHEP {\bf 9910}, 008 (1999)
[arXiv:hep-th/9909062].
}

\lref\MoellerVX{
N.~Moeller and B.~Zwiebach,
``Dynamics with infinitely many time derivatives and rolling tachyons,''
JHEP {\bf 0210}, 034 (2002)
[arXiv:hep-th/0207107].
}

\lref\gsw{
M.~B.~Green, J.~H.~Schwarz and E.~Witten,
``Superstring Theory. Vol. 1: Introduction,''}

\lref\SenNU{ A.~Sen, ``Rolling tachyon,'' JHEP {\bf 0204}, 048
(2002) [arXiv:hep-th/0203211].
}

\lref\SenIN{ A.~Sen, ``Tachyon matter,'' JHEP {\bf 0207}, 065
(2002) [arXiv:hep-th/0203265].
}

\lref\SenTM{ A.~Sen, ``Dirac-Born-Infeld Action on the Tachyon
Kink and Vortex,'' [arXiv:hep-th/0303057].
}

\lref\SenQA{ A.~Sen, ``Time and tachyon,'' [arXiv:hep-th/0209122].
}

\lref\SenAN{ A.~Sen, ``Field theory of tachyon matter,'' Mod.\
Phys.\ Lett.\ A {\bf 17}, 1797 (2002) [arXiv:hep-th/0204143].
}

\lref\SenVV{ A.~Sen, ``Time evolution in open string theory,''
JHEP {\bf 0210}, 003 (2002) [arXiv:hep-th/0207105].
}

\lref\ChenFP{ B.~Chen, M.~Li and F.~L.~Lin, ``Gravitational
radiation of rolling tachyon,'' JHEP {\bf 0211}, 050 (2002)
[arXiv:hep-th/0209222].
}

\lref\SenMS{
P.~Mukhopadhyay and A.~Sen,
``Decay of unstable D-branes with electric field,''
JHEP {\bf 0211}, 047 (2002)
[arXiv:hep-th/0208142].
}

\lref\Rey{S.~J.~Rey and S.~Sugimoto,
``Rolling Tachyon with Electric and Magnetic Fields -- T-duality approach,''
[arXiv:hep-th/0301049].
}

\lref\GutperleAI{ M.~Gutperle and A.~Strominger, ``Spacelike
branes,'' JHEP {\bf 0204}, 018 (2002) [arXiv:hep-th/0202210].
}

\lref\StromingerPC{ A.~Strominger, ``Open string creation by
S-branes,'' [arXiv:hep-th/0209090].
}

\lref\GutperleBL{M.~Gutperle and A.~Strominger,
``Timelike Boundary Liouville Theory,''
[arXiv:hep-th/0301038].
}

\lref\WaldWT{
R.~M.~Wald,
``Existence Of The S Matrix In Quantum Field Theory In Curved Space-Time,''
Annals Phys.\  {\bf 118}, 490 (1979).
}

\lref\Larsen{F.~Larsen, A.~Naqvi and S.~Terashima,
``Rolling tachyons and decaying branes,''
[arXiv:hep-th/0212248].
}
\lref\krauslarsen{
B.~Craps, P.~Kraus and F.~Larsen,
``Loop corrected tachyon condensation,''
JHEP {\bf 0106}, 062 (2001)
[arXiv:hep-th/0105227].
}

\lref\weinberg{
S. Weinberg,  ``Gravitation and Cosmology''.
}

\lref\Callan{
C.~G.~Callan, I.~R.~Klebanov, A.~W.~Ludwig and J.~M.~Maldacena,
``Exact solution of a boundary conformal field theory,''
Nucl.\ Phys.\ B {\bf 422}, 417 (1994)
[arXiv:hep-th/9402113].
}

\lref\RS{
A.~Recknagel and V.~Schomerus,
``Boundary deformation theory and moduli spaces of D-branes,''
Nucl.\ Phys.\ B {\bf 545}, 233 (1999)
[arXiv:hep-th/9811237].
}

\lref\SenMG{
A.~Sen,
``Non-BPS states and branes in string theory,''
[arXiv:hep-th/9904207].
}

\lref\juan{J.~Maldacena, Unpublished}

\lref\shiraz{S.~Minwalla and K.~Pappododimas, Unpublished}

\lref\hwang{
S.~Hwang,
``Cosets as gauge slices in SU(1,1) strings,''
Phys.\ Lett.\ B {\bf 276}, 451 (1992)
[arXiv:hep-th/9110039];
J.~M.~Evans, M.~R.~Gaberdiel and M.~J.~Perry,
``The no-ghost theorem for AdS(3) and the stringy exclusion principle,''
Nucl.\ Phys.\ B {\bf 535}, 152 (1998)
[arXiv:hep-th/9806024].
}

\lref\polchinski{
J.~Polchinski,
``String Theory. Vol. 1: An Introduction To The Bosonic String,''
}

\lref\GurarieXQ{
V.~Gurarie,
``Logarithmic operators in conformal field theory,''
Nucl.\ Phys.\ B {\bf 410}, 535 (1993)
[arXiv:hep-th/9303160].
}

\lref\fs{
W.~Fischler and L.~Susskind,
``Dilaton Tadpoles, String Condensates And Scale Invariance,''
Phys.\ Lett.\ B {\bf 171}, 383 (1986);
``Dilaton Tadpoles, String Condensates And Scale Invariance. 2,''
Phys.\ Lett.\ B {\bf 173}, 262 (1986).
}

\lref\MaloneyCK{
A.~Maloney, A.~Strominger and X.~Yin,
``S-brane thermodynamics,''
[arXiv:hep-th/0302146].
}

\lref\BuchelTJ{
A.~Buchel, P.~Langfelder and J.~Walcher,
``Does the tachyon matter?,''
Annals Phys.\  {\bf 302}, 78 (2002)
[arXiv:hep-th/0207235];
A.~Buchel and J.~Walcher,
``The tachyon does matter,''
arXiv:hep-th/0212150.
}

\lref\LeblondDB{ F.~Leblond and A.~W.~Peet,
``SD-brane gravity fields and rolling tachyons,''
[arXiv:hep-th/0303035].
}

\lref\FischlerJA{
W.~Fischler, S.~Paban and M.~Rozali,
``Collective Coordinates for D-branes,''
Phys.\ Lett.\ B {\bf 381}, 62 (1996)
[arXiv:hep-th/9604014];
``Collective coordinates in string theory,''
Phys.\ Lett.\ B {\bf 352}, 298 (1995)
[arXiv:hep-th/9503072].
}

\lref\PO{
V.~Periwal and O.~Tafjord,
``D-brane recoil,''
Phys.\ Rev.\ D {\bf 54}, 3690 (1996), arXiv:hep-th/9603156.
}

\lref\KutasovER{ D.~Kutasov and V.~Niarchos, ``Tachyon effective
actions in open string theory,'' arXiv:hep-th/0304045.
}

\lref\OkuyamaWM{ K.~Okuyama, ``Wess-Zumino term in tachyon
effective action,'' arXiv:hep-th/0304108.
}

\lref\KMW{
J.~S.~Caux, I.~I.~Kogan and A.~M.~Tsvelik,
``Logarithmic Operators and Hidden Continuous Symmetry in Critical Disordered Models,''
Nucl.\ Phys.\ B {\bf 466}, 444 (1996)
[arXiv:hep-th/9511134];
I.~I.~Kogan and N.~E.~Mavromatos,
``World-Sheet Logarithmic Operators and Target Space Symmetries in String Theory,''
Phys.\ Lett.\ B {\bf 375}, 111 (1996)
[arXiv:hep-th/9512210];
I.~I.~Kogan, N.~E.~Mavromatos and J.~F.~Wheater,
``D-brane recoil and logarithmic operators,''
Phys.\ Lett.\ B {\bf 387}, 483 (1996),
arXiv:hep-th/9606102.
}

\lref\FlohrZS{
M.~Flohr,
``Bits and pieces in logarithmic conformal field theory,''
arXiv:hep-th/0111228.
}

\lref\GaberdielTR{
M.~R.~Gaberdiel,
``An algebraic approach to logarithmic conformal field theory,''
arXiv:hep-th/0111260.
}

\lref\OkudaYD{ T.~Okuda and S.~Sugimoto, ``Coupling of rolling
tachyon to closed strings,'' Nucl.\ Phys.\ B {\bf 647}, 101 (2002)
[arXiv:hep-th/0208196].
}

\lref\Aref{
I.~Y.~Aref'eva, L.~V.~Joukovskaya and A.~S.~Koshelev,
``Time evolution in superstring field theory on non-BPS brane. I: Rolling  tachyon and energy-momentum
conservation,''
arXiv:hep-th/0301137.
}

\lref\IshidaCJ{
A.~Ishida and S.~Uehara,
``Rolling down to D-brane and tachyon matter,''
JHEP {\bf 0302}, 050 (2003)
[arXiv:hep-th/0301179].
}

\lref\KlusonAV{
J.~Kluson,
``Time dependent solution in open Bosonic string field theory,''
arXiv:hep-th/0208028;
``Exact solutions in open Bosonic string field theory and marginal  deformation in CFT,''
[arXiv:hep-th/0209255].
}

\lref\MinahanIF{
J.~A.~Minahan,
``Rolling the tachyon in super BSFT,''
JHEP {\bf 0207}, 030 (2002)
[arXiv:hep-th/0205098].
}

\lref\SugimotoFP{
S.~Sugimoto and S.~Terashima,
``Tachyon matter in boundary string field theory,''
JHEP {\bf 0207}, 025 (2002)
[arXiv:hep-th/0205085].
}

\lref\ReyXS{
S.~J.~Rey and S.~Sugimoto,
``Rolling tachyon with electric and magnetic fields: T-duality approach,''
arXiv:hep-th/0301049.
}

\lref\rastelli{ D. Gaiotto, N. Itzhaki and L. Rastelli, hep-th/0304192.}

\lref\LambertHK{
N.~D.~Lambert and I.~Sachs,
``Tachyon dynamics and the effective action approximation,''
Phys.\ Rev.\ D {\bf 67}, 026005 (2003)
[arXiv:hep-th/0208217].
}





\Title{\vbox{\baselineskip12pt \hbox{hep-th/0303139}
\hbox{RUNHETC-2003-03}
}}%
{\vbox{\centerline{Closed strings from decaying D-branes} }}

\smallskip
\centerline{Neil  Lambert$^1$,  Hong Liu$^1$ and Juan Maldacena$^2$}
\medskip

\centerline{\it $^1$ Department of Physics, Rutgers University}
\centerline{\it Piscataway, New Jersey, 08855-0849}

\smallskip

\smallskip

\centerline{\it $^2$ Institute for Advanced Study}
\centerline{\it Princeton, New Jersey, 08540}

\smallskip

\vglue .3cm

\bigskip
\noindent

We compute the emission of closed string radiation from
homogeneous rolling tachyons. For an unstable decaying D$p$-brane
the radiated energy is infinite to leading order for $p\leq 2$ and
finite for $p>2$. The closed string state produced by a decaying
brane is closely related to the state produced by D-instantons at
a critical Euclidean distance from $t=0$. In the case of a D0
brane one can cutoff this divergence so that we get a finite
energy final state which would be the state that the brane decays
into.

\Date{April 21, 2003}




\bigskip

\newsec{Introduction}

A D$p$-brane in Bosonic string theory is unstable due to the
presence of an open string tachyon in its spectrum. If we displace
the tachyon  away from the maximum it will start rolling down the
potential toward the minimum, which is the closed string
vacuum~\refs{\SenMG}. It is of great interest to understand the
``real time'' behavior of this
process~\refs{\GutperleAI\SenNU\SenIN\SenAN\SenVV\SenMS\StromingerPC\SenQA\Larsen\GutperleBL\MaloneyCK\ChenFP\Rey\MoellerVX\SugimotoFP\MinahanIF\KlusonAV\OkudaYD\Aref-\IshidaCJ}.
 For one
thing, exact time-dependent solutions in string theory are hard to
come by and studying them should yield important insights into the
fundamental structure of the theory. We would also like to
understand the final state an unstable D-brane decays into.

Analysis by Sen at the classical level in open string theory  has yielded
some interesting surprises~\refs{\SenNU,\SenVV}.
It was found that
the late time evolution of the tachyon leads to a pressureless
fluid,  called ``tachyon matter'' (see however~\MoellerVX).
Tachyon matter therefore seems to be a
new and unexpected  degree of freedom.
Hence it is natural
to understand the precise nature of tachyon matter
and in particular to take
into account the open and closed
strings created by the decay process.
In~\refs{\StromingerPC,\GutperleBL}
 quantum creation of open strings was computed and it was argued that
this effect would
destabilize the
classical solution (see also~\MaloneyCK).

In this paper we are interested in computing the creation of closed
string modes from the rolling tachyon process.
Radiation produced by the rolling tachyon has been
discussed previously  in \refs{\ChenFP,\Rey} for the massless closed
string modes. In addition supergravity solutions with rolling tachyon sources were discussed
in \refs{\BuchelTJ,\LeblondDB}. We will extend this to all massive closed string modes.
In \refs{\OkudaYD,\SenMS,\Rey} it was observed that
the ``rolling tachyon''  boundary state contains  exponentially increasing
couplings to massive closed string
modes and it was suggested in \refs{\OkudaYD,\Rey}
that these will lead to pathelogical behaviour.
However  we will see that these couplings
are not relevant for the creation of {\it on-shell physical} modes.
We find that the emission into on-shell physical closed string modes
is finite for each level. When we sum over all levels we find a
divergence in the emitted energy for $p\leq 2$ (similar conclusion
has also been reached in \refs{\shiraz}).
We also find a divergence if $p>2$ and the worldvolume of the brane
is compact with a size comparable to the string scale.
The $p=0$ case is
particularly interesting since  the computation
 suggests that all the energy of the initial brane is transfered
into closed string modes.
The $p\geq 3$ case is particularly relevant for applications
to brane world cosmological models.

It turns out that the precise closed string states that are
emitted from the brane depend on the physical interpretation of
the Sen's solution. We discuss two possible interpretations. The
first is the one implicitly advocated
in~\refs{\SenNU,\SenIN,\SenAN} where one thinks of a tachyon field
coming up from its true minimum to a point close to the local
maximum associated to the unstable D-brane and then back to its
true minimum.  The second interpretation of the solution is more
closely related to the Euclidean computations. Here we cut the
Euclidean path integral at $t=0$ and paste it to the Lorentzian
path integral. In this way the Euclidean path integral is viewed
as a prescription for setting up initial conditions at $t=0$. The
subsequent Lorentzian evolution is the one resulting from these
initial conditions. In this interpretation the Euclidean
computations are more directly related to the Lorentzian ones. In
this case the closed string radiation that comes out of a decaying
brane is basically identical to the state at $t=0$ of the
Euclidean path integral with some D-instantons added at a critical
Euclidean distance from the $t=0$ plane.

In section 2 we review Sen's description of the boundary state and
compute the closed string radiation. In section 3 we explain that
the closed string states can be viewed as being produced by
D-instantons that are sitting at a critical distance in the
Euclidean direction. Section three has some overlap with the
discussion in \rastelli .
In section 4 we describe some attempts at
computing the backreaction. We have included a few Appendices. In
Appendix A we give a second quantized description of various
formulae given in the main text.
In Appendix B we describe an effective field theory model for the
closed string creation following the discussion of \SenAN. In
Appendix C, as a warmup for the backreaction question, we discuss
a simple example of the problem of finding the deformation of a
boundary state due to a closed string deformation of the
background.

\newsec{ Closed string emission from a rolling tachyon}

\subsec{Review of Sen's computation}

The decay of a Bosonic D-brane was described in \SenNU\ as a
boundary conformal field theory which is the usual non-decaying
D-brane plus a boundary interaction of the form
 \eqn\fullbrane{
 S_{bdy} =  \int_{\sig =0} d\tau \, \tilde \lambda \,
\cosh \, X^0 (\tau)\ ,
 }
where we have set the boundary of the worldsheet at $\sigma =0$
and we have set $\alpha'=1$. This tachyon profile is interpreted as
a configuration where the tachyon comes up from
the minimum associated with the closed string vacuum,
gets closest to zero at $t=0$ and then  rolls back.
Other authors~\refs{\StromingerPC,\Larsen,\GutperleBL} have
also discussed a seemingly more realistic process where we have a boundary
interaction of the form
 \eqn\halfbrane{
 S_{bdy}^{half} = \int_{\sig =0} d\tau \, \lambda \, e^{X^0 (\tau) }\ ,
 }
that describes the unstable brane at early times which then decays.
We will call \fullbrane\ the full-S-brane and
\halfbrane\ the half-S-brane.

The boundary state for a D$p$-brane with boundary
interaction \fullbrane\halfbrane\ takes the form
 \eqn\defbn{
  |B\rangle= \NN_p  \, |B\rangle_{X^0}
 \otimes |B\rangle_{\vec{X}} \otimes |B\rangle_{bc}\ ,
 }
where the normalization constant
\eqn\Ndef{
\NN_p = \pi^{11\ov 2}(2\pi)^{6-p}\ ,
}
is the same as that for a non-decaying  D$p$-brane (corresponding to
$\lambda=0$).
$|B\rangle_{\vec{X}}$ and $ |B\rangle_{bc}$ are the usual
boundary states for the spatial and ghost part of a flat D$p$-brane.
$|B\rangle_{X^0}$, which describes the dynamics of the
rolling tachyon, has the form
\eqn\timpa{
|B\rangle_{X^0} = \rho (X^0) \ket{0} + \sig (X^0) \,
\al_{-1}^0 \tilde \al_{-1}^0
\ket{0} + \cdots
}
where $\cdots$ denotes  higher oscillation modes. The functions
$\rho, \sig$ are given by~\refs{\SenNU,\Larsen}
 \eqn\rhofull{\eqalign{
  \rho_{full}(t) & ={ 1 \over 1 + \hat \lambda e^{t} } 
  + {1 \over 1 + \hat \lambda e^{-t} } -1  ~,~~~~~~ \hat \lambda =
 \sin \pi \lambda \cr
  \sig_{full} (t) &= \cos 2 \pi \lam  +1 - \rho_{full} (t) \cr
 }}
 \eqn\rhohalf{\eqalign{
  \rho_{half}(t) & = { 1 \over 1 + \hat \lambda e^{t} } ~,~~~~~~~
\hat \lambda = 2 \pi \lambda \cr
  \sig_{half} (t) &= 2 - \rho_{half} (t) \cr
 }}
for the full-brane and half-brane respectively.

Some essential aspects of the boundary state \defbn\ can be summarized
as follows~\refs{\SenNU,\SenIN,\Larsen,\GutperleBL}:

\item{1.} If $\lam < 0$ the
system becomes singular at a finite value of $t$,
as can be seen  from \rhofull-\rhohalf.
This is interpreted as due to the fact that the potential
is unbounded from below in Bosonic string theory and the tachyon can
run off on the ``wrong side'', {\it i.e.} away from the closed string
vacuum.
This does not occur in the more realistic case of the superstring.
In this paper we restrict to $\lam > 0$.

\item{2.} For the half-brane $\lam$ is not a parameter, it can be set
to $1$ by a time translation. The full-brane system is periodic in
$\lam$ and  it can be restricted to lie between $0 \leq \lam \leq \ha$.
The half-brane can be obtained from the full-brane as a limit of $\lam \to 0$
along with a time translation.

\item{3.} For the full-brane,  
at $\lam = \ha$, the boundary
state vanishes identically for real values of time. This can be
interpreted as saying that
the system sits at the closed string vacuum. For $\lam$ lying between $0$ and
$\ha$, the larger the value, the further away the turning point is from
the top of the tachyon potential.

\item{4.} During the rolling process the energy is conserved and the stress
tensor of the system can be written as
\eqn\stre{
T_{00}  = \ha T_p (\rho (t) + \sig (t))\ ,
\qquad
T_{ij} = - T_p \rho (t) \delta_{ij}\ ,
}
where $i,j$ denote spatial directions of the worldvolume and all
other components of the stress tensor vanish. In \stre\ $T_p$ is the
tension of a Dp-brane and $\rho, \sig$ are given by \rhofull-\rhohalf.
It follows from \rhofull-\rhohalf\ that as $t \to \infty$, $T_{ij} \to 0$,
{\it i.e.} the system becomes pressureless.

\item{5.} The formula for $\rho(t)$ can be obtained\foot{Note the
relation between $\lam$ and $T_{00}$ in this effective field
theory is different from that in string theory.} from an effective
field theory model~\refs{\GarousiTR,\SenMD} for the tachyon with
the action $ S = - T_p \int V(T) \sqrt{ - \det (\eta_{\mu \nu} +
\partial_\mu T \partial_\nu T)} $ with the potential
 \eqn\repot{
 V(T) ={1 \ov \cosh{T \ov 2}} \ ,}
as we detail in appendix B, see also
\refs{\BuchelTJ,\LeblondDB,\KutasovER,\OkuyamaWM}.

\subsec{Closed string radiation}

A D-brane acts as a source for  closed string modes. With a
rolling tachyon  the  D-brane becomes a {\it time-dependent}
source. Thus generically there will be closed string creation from the rolling
process. If we ignore the interactions between closed strings,
the question then reduces to the familiar problem of free particle production
by a time-dependent source. The created state is a coherent state.
Schematically, if $a^\dagger$ is the creation operator for
a {\it physical} closed string mode, then the closed string state at late times
is proportional to
$ e^{ \alpha a^\dagger } |0 \rangle $ where $\alpha$ is essentially
the one-point amplitude of the closed string mode on the disk.
More precisely,
\eqn\ampl{
\sqrt{ 2 E} \alpha = {\cal A} = \NN_p \, \vev{V}_{disk}\ ,
}
where $E$ is the spacetime energy of the state and
$V$ is the corresponding on-shell vertex operator.
${\NN}_p$ is the normalization factor \Ndef.
Notice that this is a process which is of order $g^0$.

It is possible to show that for any physical on-shell closed
string state with {\it nonzero } energy we can choose a gauge
where there are no timelike oscillators, see \hwang . In such a
gauge the full vertex operator for an on-shell closed string state
can be chosen to be of the form
  \eqn\gauge{
  V = e^{  i E X^0} V_{sp}
  }
 where $V_{sp}$ is made out of the 25 spatial fields. This is
 equivalent to the statement that we can choose a gauge where we
 can put all the $\alpha^0$ oscillators to zero.
 It should be noted that the vertex operator $V_{sp}$ is
 constrained to be a Virasoro primary state of arbitrary conformal
 weight $\Delta$. The $L_0$ constraint gives us the value of the
 energy $E$ in terms of $\Delta$, $E = 2 \sqrt{(\Delta -1)}$.
The gauge \gauge\ is very convenient since the computation of the
one point function factorizes as
 \eqn\onepoint{
 \langle V \rangle_{disk} = \langle e^{iE X^0} \rangle_{X^0} \, \langle
 V_{sp} \rangle_{sp}
 }
 into a product of the one point function for the time part and
 a one point function for the rest.

The computation of the second factor in
 \onepoint\ is straightforward and it is
 given by the inner product of the spatial part of the boundary
 state, which is the spatial part of a standard D$p$-brane,
with  the closed string state under consideration.
The spatial part of the boundary state has the form, up to an overall
constant,
 \eqn\formofstate{
  \int d^{25-p} k  \sum_{\psi \in H_L}
   e^{i \varphi(\psi) } |\psi\rangle_L  |\psi\rangle_R |k\rangle \ .
  }
Note that the boundary state is such that we have the same
 oscillator state for  the right and left movers, up to a phase
that is irrelevant for our computations. We also have a sum over
all possible oscillator states we can make with the left movers.
So the counting of states is the same as the counting for
 an open string.
Finally, all states appearing in \formofstate\ are unit normalized.

We can separate the closed string Hilbert space into
states that are left right identical, as in \formofstate\ and
the orthogonal complement. So only states in the first subspace
will be emitted. If we take
 $V_{sp} = |\psi'\rangle_L | \psi' \rangle_R |k\rangle $
the overlap with
\formofstate\ is just a phase.

Therefore, it follows from \onepoint\ that up to a phase,  \ampl\ can be
identified with  the disk one point function
for an operator of the form $e^{i E X^0}$
\eqn\anoamp{
 {\cal A} =
 {\NN}_p \, I(E) = {\NN}_p\,  \vev{e^{i  E X^0}}_{disk} \ .
}
To evaluate \anoamp\ it is convenient to
separate the zero mode integration  with the decomposition
$X^0 = t + \hat X^0$ and we have
\eqn\timon{
I(E) = \vev{e^{i  E X^0}}_{disk} = \int_{\CC} dt \, e^{ i Et} \,
\vev{e^{i  E \hat X^0}}_{disk} \ .
}
Note that in performing the
zero mode integration one has to specify a contour $\CC$.
The computation
of $\vev{e^{i  E \hat X^0}}_{disk}$ essentially reduces (almost by definition)
to that of computing
the first term in \timpa. Thus we find that
\eqn\integralform{
I (E) = i \int_{\cal C}  dt \, \rho(t) \, e^{ i E t}
}
where $\rho(t)$ is given by  \rhofull\ or \rhohalf.
In order to determine the contour ${\cal C}$ we need some more
physical information.
The most obvious choice of contour is to take it
to run  along the real axis $ - \infty < t < + \infty$. We call
this the {\it real}  contour, ${\cal C}_{real}$.
This contour corresponds to interpreting the full-brane
as a configuration where the tachyon rises up to a value closest
to zero and then decays back again. Having the tachyon rise in this
way in the full theory involves some fine-tuning  which
we do not worry about here. For the half-brane it is necessary
to rotate $t \to t (1- i \ep)$  in order to make
the integral convergent. This corresponds to the standard choice
of vacuum for closed strings, {\it i.e.} there is no incoming radiation
from $t = - \infty$.

Since  the full-brane interaction \fullbrane\ is time symmetric around
$t=0$, there is another
interesting contour which we will call the
Hartle-Hawking  contour, ${\cal C}_{HH}$. This contour runs from
$ t =  i \infty $ along the imaginary axis
to $ t=0$ and then along the real axis to $t= + \infty$.
We can imagine the contour along the imaginary axis as preparing
the state at $t=0$. If we did this in a field theory in the bulk
the path integral up to $t=0$ along the imaginary axis would
produce a state at $t=0$. This state would be the vacuum if there
were no brane. When there is a brane the state is not the vacuum
but some other state produced by the brane. We then evolve this
state for $t > 0$ along the real axis.
Note that this prescription also creates a state  for the open
string modes on the brane. This is the  state
that we naturally get if we blindly
analytically continue the Euclidean expressions. We will elaborate more
about the physical interpretation of the Hartle-Hawking  contour
in the next section.

The integral \integralform\ for the different contours gives
\eqn\integrals{\eqalign{
i\int_{{\cal C}_{real}}  dt \, \rho_{full}(t) \, e^{ i E t}  = &
(e^{-i E \log \hat \lambda }  -  e^{i E \log \hat \lambda } ) { \pi
\over \sinh \pi E }\ ,
\cr
i\int_{{\cal C}_{real}}  dt \, \rho_{half}(t) \, e^{ i E t}  = &
 e^{- i E \log \hat \lambda } { \pi
\over \sinh \pi E }\ ,
\cr
i\int_{\cal C_{HH}}  dt \, \rho_{full}(t) \, e^{ i E t}  = &
  e^{- i E \log \hat \lambda } { \pi
\over \sinh \pi E }\ .
}}
The two exponentials in the first line of \integrals\ can
be interpreted as  the
radiation emitted in the rising  part of the full-brane and
the decaying part of the full-brane. Note
that the $\hat \lambda$ dependence for the half brane
is precisely what we expect on the basis of time
translation invariance.

Surprisingly, the result for the full-brane using the
Hartle-Hawking contour gives the same result as the half-brane.
 {} From \ampl\ and \anoamp\ this implies that, up to a phase due to time
translation, the final closed string state produced by
the half-brane is the same as that produced by the full-brane
using the Hartle-Hawking  contour.
Since $\hat \lam$ only appears  in the  phase, the closed string
states with
different values of $\hat \lambda$ only differ by an
overall time translation. {}For the half-brane this is a trivial  consequence of the fact that
changing $\lam$ is the same as performing a time translation,
 but for the full-brane with $\CC_{\HH \HH}$ this is surprising.
Note that the Hartle-Hawking presciption leads to a time symmetric
state. So we have radiation coming in from $t= -\infty$ and being
absorbed by the brane. For $ \lambda =1/2$ we have no D-brane
and this radiation that comes from $t=-\infty$ just reflects from
the origin and goes out.
 In contrast,
for the full-brane with the real contour,
the absolute value of the amplitude
is $\lam$ dependent and vanishes at $\lam = \ha$.
For $\lambda \to 0 $ we see that the closed strings  produced by the
rising phase and decaying phase is moved
all the way to $t = \mp \infty$.

Note also that the contour that is natural in the Euclidean theory,
which runs along the imaginary axis, would be badly divergent for
real $E$.
Nevertheless \integrals\ are analytic in $E$ and when $E = i n$
they have a pole. We see that the residue of the pole for the
Hartle-Hawking countour or the half-brane has the values we get
in the Euclidean computation. The fact that there is a pole
is related to the fact that the Euclidean computation for $E=in$
has a divergent volume factor, the volume of Euclidean time.
 Note that that the residue of the
pole for the real contour is {\it not} the same as the result for
the Euclidean computation. As we explained above the Euclidean
computation is more directly related to the Hartle-Hawking
contour.

In conclusion the emission amplitude for this closed string state
will be given by the integrals \integrals\ with $E$ equal to the
spacetime energy of the state.

Now we compute the total average number and the total energy of
particles emitted.
Since the final state is a coherent state, from \ampl,\anoamp\ we have
\eqn\averagen{
{\bar N \ov V_p} = \sum_{s} {1 \over 2 E_s} |{\cal A}_s|^2  =
\NN^2_p \sum_{s} {1 \over 2 E_s} |I(E_s)|^2\ ,
}
\eqn\averagee{
{\bar E \ov V_p} = \sum_{s} {1 \over 2 } |{\cal A}_s|^2  =
\NN^2_p \sum_{s} {1 \over 2} |I(E_s)|^2\ ,
}
where the sum runs over a basis of  physical left-right
identical closed string states and $V_p$ is the spatial volume of the
D$p$-brane. Note that the sum over $s$  includes both the sum over
level $n$ and
an integral over the momenta of the spatial directions transverse to
the brane.

It is interesting to estimate the  behaviour
of \averagen\ and \averagee\ for large $n$ to check if the result
is finite or not. We see that
\eqn\asymp{
|I(E)|^2 \sim e^{- 2 \pi E} ~, ~~~~~~~~~~~D(n) \sim
{1\ov \sqrt{2}}n^{- {27 \over 4}}
e^{ 4 \pi \sqrt{n} }\ ,
}
where
$D(n)$ is the number of primary closed string oscillator states
that are left-right identical at level $n$. This number goes  as the
number of states in the open string Hilbert space, see~\gsw .
Note that
the boundary state in the spatial directions
 \formofstate\ contains 25 oscillators, while
the degeneracy in \asymp\ grows as that of 24 oscillators. This
is due to the fact that the condition that the state is a Virasoro
primary eliminates precisely one oscillator worth of states. This
is due to the fact that the Virasoro characters for $c>1$ and
$\Delta >0$ have no null states so that their counting is precisely
the same as the counting of a single oscillator.

The energy is related to the level $n$ by
\eqn\energ{
 E = \sqrt{ \vec k_{\perp}^2 + 4 n} \sim 2 \sqrt{n} + {
 \vec k_{\perp}^2 \over 4 \sqrt{n} } + \cdots
} where we have expanded the result for large levels. When we
insert this in \averagen\ and \averagee\ we find that the
exponential terms are exactly cancelled. In addition the integral
over $\vec k_{\perp}$ produces factors of $n^{1/4}$ per additional
transverse dimension. The result takes the form \eqn\NEsum{ {\bar
N \ov V_p} \sim \sum_n n^{-{p \ov 4}-1} ~,~~~~~~~~~~~ {\bar E \ov
V_p} \sim \sum_n  n^{-{p \ov 4}-{1\ov 2}}\ . } Note that the
emitted energy is finite for $p>2$ and infinite for $p\leq 2$. In
the cases that the expression for the energy is finite we find
that the total energy emitted is of order $g^0$ (in $\alpha'=1$
units). In the divergent cases the sum is dominated by the large
$n$ terms and \averagen, \averagee\ can be written as
 \eqn\averages{\eqalign{
 {\bar N \ov V_p} & = (2 \pi)^{-p}  \int
 {dn \over 2 n} (4 n)^{-{p \ov 4}}+  \cdots \cr
 {\bar E \ov V_p} & = (2 \pi)^{-p} \int d E  E^{-{p \ov
 2}} + \cdots \ , \cr 
 } }
where we restored the correct numerical factors.

Of particular interest is the $p=0$  case. In this case the number
of emitted particles diverges logarithmically and the emitted
energy diverges linearly. If we cutoff the energy at $E \sim 1/g$
we see that we get an energy of the same order of magnitude as
the initial
mass of the brane. Of course the energy emitted cannot be greater
than the mass of the initial brane, so higher order corrections
must cut it off so that the total emitted energy is finite.
What we find suggestive is that energies of order $E \sim 1/g$ are
precisely the energies where we expect the tree level computation
to become invalid, since at this energy the gravitational force
between the emitted state and the original D-brane become of order 1.

\subsec{Remarks on the fate of branes}

The above computation strongly suggests that a D0-brane decays
completely into closed
strings.
Note that according to  the formulas \averages\ most of the energy
is emitted into very massive string states. These massive string
states are roughly the same as the ones that appear in the boundary
state, so they represent closed strings highly localized at $r=0$
in the transverse directions.
Since their momentum is of order $n^{1 \ov 4}$
and mass of order $n^\ha$, they move very slowly at the speed of
$n^{-{1 \ov 4}}$ away from the origin.
Thus the time scale for having some
matter localized within string scale around the origin is
of order $g^{-\ha}$.

 For a D$p$-brane ($p
> 2$) with an noncompact worldvolume (or in a very large compact
space) equations in \averages\ appear to suggest that the amount
of energy going into closed string modes is rather small and one
might conclude that the tachyon matter is stable at least in
perturbation theory\foot{This paragraph was added in Feb. 2007
thanks to comments by J. Polchinski and it improves our original
discussion.}. In order to understand this more precisely
  it is useful to consider the case where we compactify
the Dp brane on a very large volume. In that case, we could also
consider the decay into winding modes. If we also include these
winding modes, then the problem is similar to considering a D0
brane decaying on a very small compact space. For a D0 brane the
mean value of the transverse momentum for a string at level $n$ is
of order $ k_\perp \sim n^{ 1/4} $. For a D0 brane we expect that
$\sqrt{n}$ is of the order of the initial energy of the D0 brane,
$1/g_0 \sim \sqrt{n}$. Thus, most of the energy will come out in
states which have a momentum of order $k_\perp \sim { 1 \over
\sqrt{g_0} }$. In order for this computation to be well defined we
need that this momentum is much bigger than the inverse size of
the space. If we started originally with a Dp brane on a torus
with all circles of radius $R_p$, then we find that the associated
D0 brane lives on a space of size $R_0 = 1/R_p$ and the string
coupling is $g_0 = g_p/(R_p)^p$. Thus, we see that for small fixed
coupling $g_p$ and very large sizes $R_p$, then we always find
that, for $p\geq 2$, the mean momentum (in the D0) brane picture
is $ k_\perp \sim { (R_p)^{p\over 2} \over \sqrt{g_p}  }  \gg { 1
\over R_0} = R_p $. Note that $g_0$ is also very small in the
large $R_p$ limit. When we go to the original Dp brane picture
this is saying two things. First, it is saying that the total
energy is still divergent if we include winding modes. Second,
that most of the energy is coming out in winding modes. These are
strings that are wound along the large directions of the torus.
These strings still give the dominant contribution to the energy
even in the infinite volume limit, where these strings are
infinitely long.

We would like to emphasize that the homogeneous decay as in
\fullbrane\ and \halfbrane\ is hardly physical for $p \geq 1$. In
these cases, long wavelength tachyonic modes with $k^2 < 1$ are
also unstable and will grow in time  \refs{\SenVV,\Larsen}.
 This means that the D-brane
decay will be rather inhomogeneous. The energy density of the
D-brane will tend to concentrate around the minima of the
potential and the original D$p$-brane may be thought of
decomposing into a collection of small patches around various
minima, see \refs{\SenVV,\Larsen}.

\subsec{Superstring case}

The boundary state for an unstable D-brane in   superstring theory
 is rather similar to the
bosonic one and was discussed in~\refs{\SenIN,\Larsen}.
The basic results are also rather similar. Indeed the profiles $\rho_{full}$ and
$\rho_{half}$ are of the same form. The on-shell tachyon
has the form
$T = \lambda e^{\pm X^0/\sqrt{2}}$ and we get an extra square root of
two in the exponents for the profiles \rhofull \rhohalf .
  Therefore the
Fourier transforms have a weaker exponential fall off as a function of the
energy. However the exponential growth in the density of states
for the superstring is also slightly slower and again the two cancel
so that $\bar N$ and $\bar E$ have exactly the same form as \NEsum.
Hence the emitted energy also diverges
linearly with energy for a decaying unstable D0-brane.
Note that in the bosonic string the inhomogenous decay is potentially
ill-defined since tachyon can also be negative. In particular
one might worry that small fluctuations near the top of the tachyon
potential could send parts of the D$p$-brane towards the vacuum with
unbounded energy, rather that the closed string vacuum. For superstrings
this is not a problem since the potential is bounded from below.

\subsec{One loop diagram}

In the previous subsection we computed the closed string radiation in
a rather direct way by computing the one point functions that
can be computed from a variety of
methods~\refs{\SenNU,\Larsen,\GutperleBL}.

The D-brane state acts as a classical source for closed strings.
So the one loop Feynman amplitude would give us information about
the radiation that comes out. Equations \averagen\averagee\ can be precisely
understood this way and we give a derivation based on the second quantized
theory in Appendix~A.
The formal expression for the one loop
amplitude in the closed string channel is
 \eqn\atr{
 W = \ha  \biggl \langle B \biggl |
 {b_0^+ c_0^- \ov L_0 + \bar L_0 - i
 \ep} \biggr | B \biggr \rangle \ .
  }
One can show that the imaginary part
of the above one loop effective action is related to the total number
of particles radiated, {\it i.e.}
\eqn\defbaN{
 \bar N
   = 2 {\rm Im} W =      {\rm Im} \,
 \biggl \langle B \biggl |
 { \, b_0^+  c_0^-  \ov L_0 + \bar L_0 - i
 \ep} \biggr | B \biggr \rangle \ .
 }
Similarly the energy emitted can be written as
\eqn\deftoen{
\bar E =
{\delta \ov \delta a}  \biggl \langle B \biggl |
\left({\, b_0^+  c_0^- \ov L_0 + \bar L_0}\right)_{ret}  \biggr
| B (a) \biggr \rangle \biggr|_{a=0}\ ,
}
where $\ket{B(a)}$ is obtained from $\ket{B}$ by taking $t \to t+a$, i.e.
they are separated by an interval $a$ in time direction. The subscript ``ret''
indicates that one needs to use the retarded propagator. One can check that,
level by level, \defbaN\ and \deftoen\ agree with \averagen\ and \averagee.

For a static D-brane, \atr\ is related by modular invariance to the
one-loop partition sum of the open string theory. In our case, such a relation
is  not clear
since we have a Lorentzian cylinder. However, one can show that the imaginary
part \defbaN\ does have an interesting
 open string interpretation. More explicitly,
one finds that after summing over all levels, for the
half-brane, \defbaN\ can be written in the open string form
 \eqn\totn{
 {\bar N \ov V_p} = \sum_{n=1}^\infty n \, \int_0^\infty {ds \ov s} \,
 {1 \ov (4 \pi \apr s)^{p \ov 2}} \, e^{-n^2 s} \, {1 \ov \eta^{24}
 \left({i s \ov 2 \pi}\right) } \,  .
 }
The above equation can also be rewritten as the following
 \eqn\rewbo{
 {\bar N \ov V_p} = \vev{B_- \biggl|{b_0^+ c_0^-  \ov L_0 + \bar L_0 } \biggr|
 B_+} \ ,
 }
where $\ket{B_{\pm}}$ are defined by
\eqn\arraydinst{
\ket{B_{+}} = \sum_{n=0}^\infty\ket{B_p [x=(2n+1) \pi ]}
  ~,~~~~~~~~~~~~~~~
\ket{B_{-}} = \sum_{n=0}^\infty\ket{B_p [x=-(2n+1) \pi ]}\ ,
}
with $\ket{B_p [x]}$ the boundary state of a {\it Euclidean}
D-instanton
with $p$ spatial dimensions sitting
at position $x$ in
the Euclidean time direction.
We see that \rewbo\ includes only open strings which stretch across the
$t=0$ line. This is different than the naive Euclidean computation
for an array of D-instantons (corresponding to $\lambda =1/2$ for
the full-brane), which would include open strings stretching between
any two D-instantons. Note that we get the result \rewbo\ for the
half-brane or the full-brane with Hartle-Hawking contour
independently of the value of $\lambda$.
 Note also that the overlaps of boundary states include
both orientations of stretched open strings.
 The corresponding formula
for the full-brane with the real contour
 is somewhat more complicated, and can be obtained by
replacing $e^{-n^2 s} $ in \totn\ by
\eqn\fullop{
2 e^{-n^2 s} - \exp
 \left[\left({\log \hat \lam \ov \pi} - i n\right)^2 s\right]
- \exp \left[\left({\log \hat \lam \ov \pi} + i n\right)^2 s\right]\ .
}
It is not clear whether the
above equation has an open string interpretation.
The expression
\fullop\  appears to come from
 D-instantons sitting at complex positions in the
complex $t$ plane.

Another way to derive ~\rewbo\ is as follows.
Note that for the half-brane we can
rewrite \anoamp\ and \integrals\ as
 \eqn\euab{\eqalign{
 {\cal A} & = N_p I (E) = N_p \vev{e^{i E
 X^0}}_{\oint \lam e^{X^0}} \cr
 & =  2 \pi N_p \, e^{-i E \log \hat \lam}
 \sum_{m=0}^\infty e^{- \pi (2m +1) E} \cr
 & = e^{-i E \log \hat \lam} \, \sum_{m=0}^\infty N_{p-1} \vev{e^{i E
 X^0}}_{X^0 = i (2 m +1) \pi}
 }}
where in the last line above $\vev{}$ denotes one point function
for a D-instanton sitting at a point on the imaginary time axis.
Thus when we sum over all intermediate closed string modes in
\averagen\ we find \rewbo. The appearance of D-instantons can be
understood as follows. In obtaining $\rho(t)$ \rhohalf\ in
\integralform\ we  treat the worldsheet boundary interaction
\halfbrane\ in terms of perturbative expansion
$$\sum_n {1 \ov n!} \left(\oint d \tau \, \lam e^{X^0}\right)^n$$
However, the perturbation series only have a finite radius of
convergence and this is reflected in the singularities of $\rho
(t)$ on the complex plane. For example, $\rho (t)$ in \rhohalf\
has poles at $t = i (2m +1) \pi + \log \hat \lam$. The
contributions of these poles to the integrals \integrals\ are
precisely given by the D-instantons discussed in \euab.
When there are poles on real axis, as
is the case for $\hat \lam < 0$ in \rhofull, \rhohalf\ and some
inhomogeneous decay examples~\refs{\SenVV,\Larsen}, \integralform\
is not unambiguously defined.

Equation \totn\ gives a very clean picture of the divergence in $\bar N$
for $p=0$ in terms of the open string picture.
This is an {\it IR divergence} (arising at large $s$)
that comes from
the open string stretching between the
two closest D-instantons sitting at $x= \pm \pi$.
The distance between the two D-instantons is such that this open string
is precisely massless. This gives rise to the divergence.

The expression for the energy can be obtained by considering
\eqn\expren{
 { \bar E \over V_p} = \left. {\partial \over \partial a }
\vev{B_- \biggl|{b_0^+ c_0^- \ov L_0 + \bar L_0 } \biggr|
 B_+(a) } \right|_{a=0}  \ ,
 }
where $ \ket{B_+(a)}$ corresponds to the same array
we had in \arraydinst\
but shifted by $a$ in the Euclidean time direction. In other words,
the open strings that had masses squared proportional to
$ (2 \pi n)^2-1 $
have masses proportional to $ (2 \pi n + a)^2 -1 $ after we shift
$\ket{B_+}$ by $a$.

If $0<p \leq 2$ we get a divergence in the energy which is
related  to the fact that a massless field (the massless
 open string mode
stretching between the two D-instantons closest to $t=0$) does
not have a well defined vacuum in two or less Euclidean dimensions.


Note that if we treated the closed string theory as an ordinary
field theory
we would have to sum over real Lorentzian
 momenta in the time direction.
 In other words the one loop diagram becomes
\eqn\oneloop{
{ \cal Z}_1  =
 \NN_p^2 \sum_n  \int { dk \over 2 \pi} { 1 \over -k^2 + E_n^2 - i \epsilon}
I(k) I(-k)\ ,
}
where the sum is over all left-right identical closed string states
with energies $E_n$. As above the sum over $n$ includes an integral
over the transverse directions. It is easy to see that, level by
level, the imaginary part of \oneloop\ indeed agrees with
the expression for $\bar N$ \averagen .
The normalization factor can be computed by noticing that for
$\lambda \to 0 $ we should recover the usual expression for
Bosonic D-branes \polchinski .

Note that the expression \oneloop\ does not include possible
$\delta(k)$ contributions which according to Sen, \SenNU ,
 are present in the
boundary state. One example is the contribution that gives
rise to the conserved energy.
The Euclidean boundary state also contains contributions proportional
to $\delta(k-in)$, it is not clear if these should be included or
not.

In~\refs{\StromingerPC,\GutperleBL}, open string creation was
discussed. One would expect this to be related by modular
invariance to the computations done in this paper. We could not
make this relation explicit. The amount of energy that was lost
into open string modes, according
to~\refs{\StromingerPC,\GutperleBL}, grows exponentially with
time,  with a coefficient that is finite for $p<25$ and infinite
for $p=25$. Computations in the open string description seem harder
because we do not expect to have asympotic
physical open string states once
the tachyon condenses. On the other hand, closed string states are
still well defined in the future.

\ifig\unstable{
Lorentzian picture for the Hartle-Hawking state for the full brane.
The state is time reflection symmetric. We have incoming closed string
radiation from the past and outgoing in the future. It is indicated by
arrows.
In a) we show the state for $\lambda =1/2$ which contains only incoming
and outgoing radiation. In b) we show the state for $\lambda < 1/2 $,
the incoming radiation is absorbed by the brane and then reemitted.
}
{\epsfxsize3.5in\epsfbox{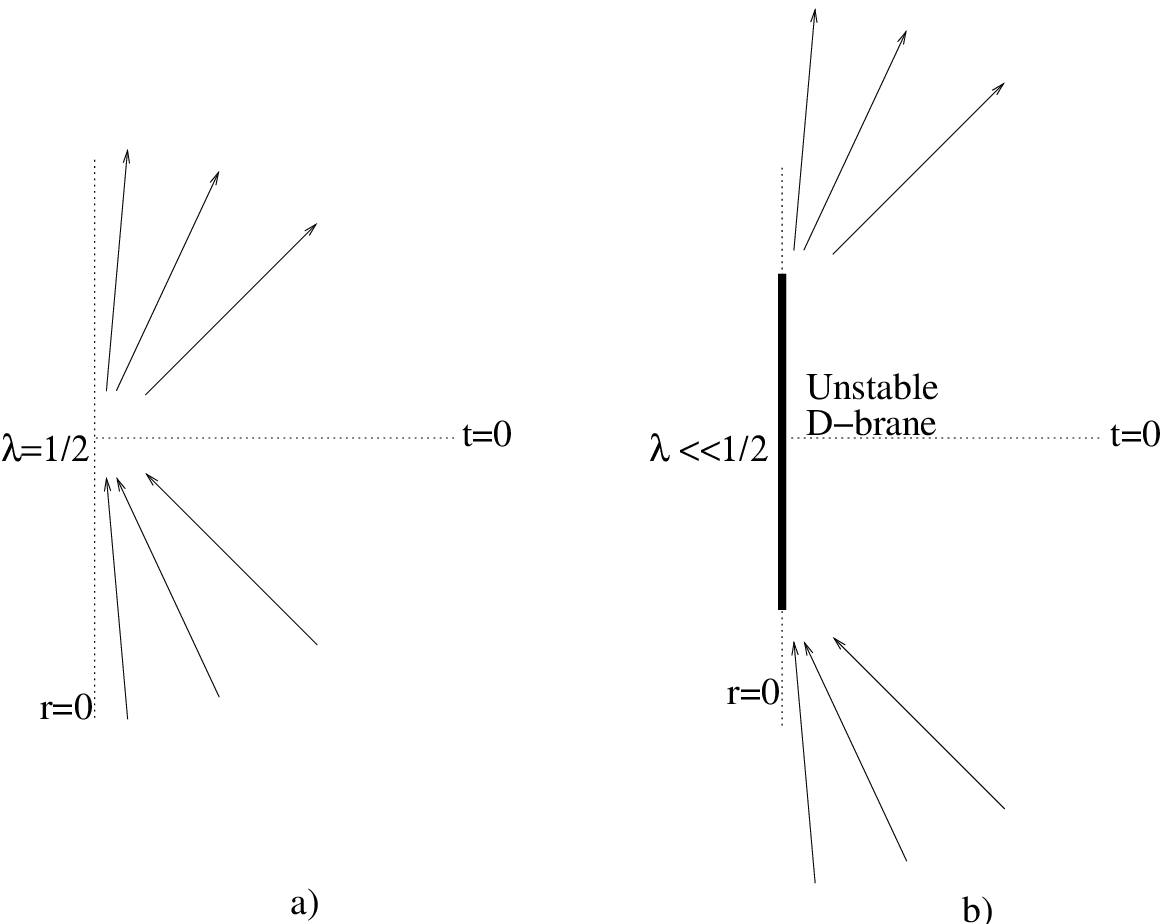}}

Note that for the full brane with the Hartle-Hawking contour the
state that is emitted depends on $\lambda$ only through a simple
phase. So the only difference in the state emitted is an overall
time translation. Since the state is time reflection symmetric
the incoming closed strings are also translated by a similar amount
into negative times. This is schematically represented in
 \unstable .
If we assume that  the  D0-brane  decays completely into these
closed string modes, then we get an intriguing picture where
finely tuned incoming closed strings  produce
 an unstable D0 which after a while decays again into closed strings.

\newsec{D-instantons and the closed string final state}

There is a very beautiful and simple description
of the closed string state that is produced by the full-brane
with the Hartle-Hawking contour or the half-brane\foot{This discussion
has some overlap with the discussion in \rastelli , where these
issues are discussed in more detail.}.
 As we explained
above the state that is produced is a coherent state in the space
of closed string fields. Let us think for the moment about the
bulk closed string field theory as an ordinary field theory.
A method for producing coherent states is to perform the path integral
in Euclidean space with a source
and cut this path integral at $t=0$. The result
of the path integral is a function of the boundary
values of the fields at
$t=0$. We can think of this function as a wavefunction for the fields.
For example, the wavefunction of the vacuum is obtained by doing the
path integral over negative  imaginary Euclidean time.
If we insert operators or sources
at Euclidean times $t_{E} <0$ this same
path integral will  produce a different state.

For simpliticy let us start by discussing the state that we get if
we insert a D-instanton at a distance $d$ from $t=0$ (see fig. 2).
This will act as a point source for many closed string fields
which at $t=0$ will have an amplitude of the form
$e^{ - d E}$ where $E$ is the energy of the particular closed
string mode. The norm of the state involves squaring the
amplitude and summing over all states. The terms in this
 sum  go as $e^{ - 2 d E} e^{ 2 \pi E} $ for large $E$. So
see that if $d < \pi$ the state that we produce in this way is not
normalizable.
Therefore the approximation in which we derived
it is not valid and we do not trust the resulting state.
On the other hand if $d > \pi$, the state that we create in this
fashion is a perfectly nice and normalizable finite energy state
in the field theory of closed string modes and has an energy of order
one ({\it i.e.} $g^0$).
We can understand this change in behaviour as follows.
We can compute the norm of the state produced in the above way by
considering the path integral over  full Euclidean time were
we put sources at $t>0$ in a time reflection symmetric fashion.
The divergence in the norm is related to the
fact that the open string going between the D-instanton at $t= -d$ and
the one at $t = + d$ becomes tachyonic precisely at $d= \pi$.
So this divergence is  rather similar to
the Hagedorn divergence. Note, however, that the state that is produced
in this fashion is not thermal, it is a pure state.

\ifig\instantons{
Construction of a closed string state via the insertion of D-instantons
in Euclidean space. In a) we pictorially represent doing the path integral
over half of the Euclidean time line, $t_E <0$,
 with an instanton inserted  and
in b) we represent the computation of the norm of the state. This norm
becomes ill defined if the D instanton distance to $t=0$
is $d< \pi$.
}
{\epsfxsize3.5in\epsfbox{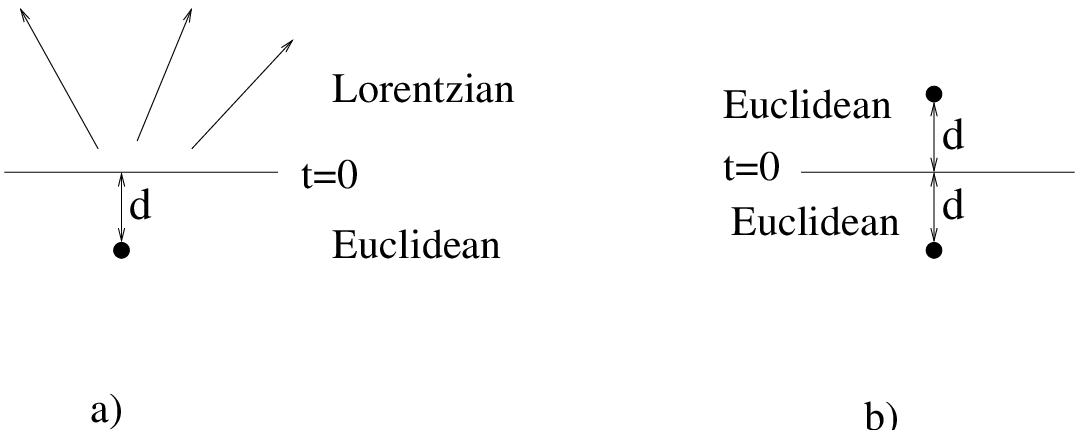}}

Now, instead of a single D-instanton we can consider an array of
D-instantons sitting at $t = - i \pi ( 1 + 2 m)\ , \
m=0,1,2,\cdots $. The large energy behaviour of the amplitudes is
dominated by the D-instanton that is closest to $t=0$. It is
interesting that the states arising during tachyon decay
correspond to D-branes at $d = \pi$ in Euclidean space. When we
compute the norm of the state we have the array of D instantons at
$t_E = i \pi ( 1 + 2 m)$, $m \in Z$. This is precisely the
Euclidean state for $\lambda =1/2$, for which
there is no brane in Lorentzian signature \SenNU . However, when
we computed the closed string radiation in \integrals\ we found
some radiation coming out if we used the Hartle-Hawking contour.
But this radiation is the same radiation that is coming in from $t
= -\infty$ in a time symmetric fashion, see \unstable .
The state at $t=0$
contains no brane because the D-intantons are localized in
Euclidean time away from $t=0$. Note that this description is
completely explicit. For example we can consider the dilaton
field. Its expectation value in
 Euclidean space is given by the expression
(see also \MaloneyCK)
 \eqn\dilfield{ \phi(r,t_E) =  \sum_{m=-\infty}^\infty
{ 1 \over  [ r^2 + (t_E - \pi (1 + 2 m))^2]^{24-p \over 2} }\ .
 }
which in Lorentzian signature gives
 \eqn\dilfieldlor{ \phi(r,t_L) =
 \sum_{m=-\infty}^\infty
 { 1 \over  [ r^2 + (i t_L - \pi (1 + 2 m))^2]^{24-p \over
 2} }\ .
 }
If we compute the positive frequency components of \dilfieldlor\
we find again an expression like \integrals\ with $E = |k|$ where
$k$ is the momentum in the directions transverse to the
D-instanton used to produce \dilfield . This  D-instanton has $p$
spatial dimensions. As expected \dilfieldlor\ obeys the massless,
sourceless field equation in the bulk. It represents a dilaton
wave coming in to $r=0$ and then going back to $r \to \infty$ (see
\unstable).

Our previous formulas describe completely the coherent state that
is produced by this process. Just to clarify the situation a bit more
let us point out that the closed string state that is produced by
a D-instanton sitting at position $d$ in Euclidean space is given by
\eqn\closed{
 | C \rangle  = P_{phys} e^{ -d \hat E} |B\rangle
}
where $|B\rangle$ is the boundary state for a D-instanton at the origin.
$\hat E$ is the operator that measures the energy of the state. And
finally $P_{phys}$ is a projector on to physical states. The state
$|C \rangle$ is a state in the closed string Hilbert space as long
as $d>\pi $. The actual state in the full multiparticle closed string
Hilbert space is the coherent state associated to the
state \closed . In other words, if we call $\alpha a^\dagger$ the
operator
creating a single string in the state \closed , then the full state
is $e^{\alpha a^\dagger} |0\rangle $.

For the superstring, at $\lambda =1/2$,
 the corresponding Euclidean configuration
involves an array of alternating
D-instantons and anti-D-instantons. This is due to the fact that
the RR fields get an $i$ when we continue to Euclidean signature so
they need to get a minus sign when we perform a time reflection in
Euclidean signature in order to be real in Lorentzian signature. The
alternating charges ensure that.

Note that these instanton produced
states are in the spirit of the  S-branes discussed in
\GutperleAI\ but differ from the D-s-branes localized in
lorentzian time that
were discussed in \GutperleAI\  by the crucial fact that
they are displaced by a critical  distance in Euclidean space.
This displacement
makes sure that the state that is produced does not have a badly
divergent norm.
If we go back to Lorentzian signature the closed string
state that is created this
way is time reflection symmetric around the origin. Note also that the
state we produce  is not $SO(26-p)$ invariant in Euclidean space, so
it does not have $SO(1,25-p)$ symmetry in Lorentzian signature.

There is one interesting aspect that makes the string theory computation
different from a field theory computation. In a field theory we are
free to add any sources we want when we do the Euclidean path integral.
In string theory it is not clear that one could do that since one
does not have local operators. The D-instantons are objects in the
theory that in the classical limit act as heavy fixed sources for the
closed string fields. But as we go to higher orders the D-instantons
become dynamical and can, for example, move.
It would be nice to understand more precisely how to view
the setup that we have been describing in a more
non-perturbative way. One would have to find a natural context in
which one can produce these states. One possibility is to
consider the thermal partition function, for temperatures
$\beta  = 2 \pi n$ (we need $n>2$ in order to be at temperatures lower
than the Hagedorn temperature). See \MaloneyCK\ for further discussion.
  This thermal partition function contains
contributions from the Euclidean branes with arbitrary $\lambda$
(in fact we should integrate over $\lambda$ when we compute the
thermal partition function). We could then imagine that we somehow
analytically
continue these to Lorentzian signature. So that we interpret the
process as the spontaneous nucleation of such a brane in the thermal
background which then subsequently decays.

In \MaloneyCK\ a  different interpretation was proposed
for the $\lambda =1/2$ state. This difference is sharper for
$p>2$. In \MaloneyCK\ it was proposed that such a state carries
an energy density
of order $1/g$. Here we are saying that this state carries
an energy density of order 1. For $p=0$ we find a divergent energy and in
fact we expect that this divergence is cutoff at an energy of
order $1/g$. \foot{ In fact, the discussion in \MaloneyCK\ could
 also be carried out with a D-instanton array where the instantons
 are separated by a distance bigger than the critical distance. In that
 case our description would give, in the Lorentzian theory,
 a state of energy of order $g^0$ while
 a computation in the spirit of \MaloneyCK\ would suggest a state with
 energy
 proportional to $1/g$.
We think that the discrepancy is related to the different translation
between the Euclidean and Lorentzian computation.}

\newsec{ Attempts of computing the back reaction}

For a D0-brane we observed that the one loop diagram
diverges.\foot{ For higher D$p$-branes there are divergences in
the expectation values of the energy, or higher powers of the energy.
Since these are measurable observables that should give finite answers,
we expect corrections for these too. }
When such a thing happens in string theory, one is supposed to change
the classical string solution in order to cancel the divergence \fs .
In the examples discussed in \fs\ the divergence comes from the IR
of the closed string channel and consequently one needs to change
the IR properties of the closed string background.
In our case the divergence comes from the UV of the closed string
channel, or the IR of the open string channel. So we conclude that
the divergence should be eliminated by modifying the long time
behaviour of the open string classical solution.

We know that Sen's state obeys the classical open string field theory
equations.
In principle one would like to solve the one-loop corrected
open string equations of motion.
This seems to be a rather difficult task. Some of  the difficulties
were described in a related context in \krauslarsen . These
difficulties come form the fact that we need to evaluate both the disk
and the cylinder diagram off shell.
 However we are only interested
in the changes of the open string solution over relatively
long time scales.
One could object that once one tries to balance classical equations
with one loop corrections one needs to include all loops. What makes
the situation a little more hopeful in our case is that the emission
of very massive closed string modes
might be a relatively slow process and could change
the boundary state rather slowly.

In any case, we did not succeed in finding the corrected equations
and we will just present a summary of our attempts. Hopefully a
more satisfactory solution will be found soon.

Our idea was that the parameter $\lambda$ which appears in the
boundary state will no longer be a constant but it would
slowly evolve in time\foot{
More precisely we write the boundary interaction on the worlsheet
as $S = \int \lambda_+e^t + \lambda_- e^{-t} $ and assume that
$\lambda_\pm$ depend on time.}.
 This would have two effects. One would
be to provide an effective cutoff for the energy of the emitted
radiation. The second is that the boundary state would slowly
approach $\lambda =1/2$ for the full-brane which corresponds to
having no D-brane, only outgoing radiation.
We did not succeed in showing that this is indeed what happens
but we will describe the difficulties we ran into.

A basic simple example to keep in mind is the following. Consider
a heavy degree of freedom, localized at the origin which couples
to a free field that propagates in the bulk. The heavy object
is moving along a solution of its classical action. The coupling
to the free field would imply that some radiation is produced and
this will suck energy away from this degree of freedom.
If the action for the total system has the form
\eqn\totaact{
S = S[q(t)] + \int dt j(q(t)) \phi(t, x=0) + {1 \over 2} \int dt d^d x
 (\partial \phi)^2 - m^2 \phi^2
}
Then the corrected equations of motion for the particle are
\eqn\eqmot{
{\partial S \ov \partial q(t) } +  { \partial j \ov \partial q(t) } \int
dt'  G_{ret}(t,0;t',0) j(q(t'))=0\ ,
}
where $G$ is the retarded propagator, it is evaluated at the origin
in the transverse dimensions.
We would like to find an expression which is equivalent to this
in string theory. In principle one could do this in string field theory.
In boundary string field theory equations of this type were
contemplated in
\krauslarsen\ in a similar context.

By taking test functions and integrating them over time against \eqmot\
we can find a particular subset of equations.
A simple one is the energy conservation condition, which arises
when we multiply by $\dot q(t)$ and integrate \eqmot . The resulting
equation says that the energy lost by the system is equal to the energy
emitted in radiation.
So this equation will already be enough to make sure that energy
is conserved. In order to get this equation to work in string theory
we only need to assume that the classical energy  vanishes at late times
and that it is proportional to $ - \int M_0$ for early times. Then
the equation \eqmot\ implies that
\eqn\eqmotst{
 M_0 = \bar E \ .
}
But this is rather trivial, what we really need to show is that
at late times the D-brane state has zero energy.

A simpler problem than the one we have is the following. Suppose
you have  D-brane in flat space. Now you deform flat space by
changing the closed string background. This change is described
by adding the operator ${\cal O}(z,\bar z)$ on the string worldsheet.
The old boundary state that described the D-brane in flat space is
not a good conformal boundary state in this perturbed bulk CFT.
In particular, when ${\cal O} $ approaches the boundary described
by the old boundary condition we will find an operator product
expansion of the form
\eqn\opeexp{
{ \cal O}(z,\bar z) \sim  \sum_n (z-\bar z)^{ -2 + n} { V_n}
}
where $V_n$ are boundary operators of conformal weight $n$.
The term proportional to $V_1$ will give rise to a logarithmic
divergence when we integrate over $z-\bar z$.
This logarithmic divergence is cancelled by adding to the
boundary an interaction of the form
\eqn\boundaddit{
\int d\tau V_B(\tau) ~,~~~~~~~~~~ {\rm with} ~~~(L_0 -1) V_B = V_{1}\ .
}

In appendix C we show how this works for a simple example.
In general  one  would think that we also need to consider the
operators $V_n$ with $n \not =0$.
However, the operator
$V_1$ seems  the  most important one at long
distances in spacetime.

Now going back to our problem. When a closed string mode is emitted,
the boundary state now moves in the background generated by
 this closed string
mode. So one would attempt to find the correction to the
boundary state by performing an analysis like the one above for each
closed string state and then
summing over all states.

Let us consider first the half-brane. In this case the interesting
term in the OPE is
\eqn\euclope{
e^{ i n X }  \sim { 1\over (z -\bar z)^{n^2 -1} } \lambda^{ n}
 n ( { J^- \over \lambda }  + {1 \over 3} \lambda J^+ ) \ .
}
when we compute it in the Euclidean theory. $J^3 = i \partial X$ and
$J^\pm =e^{ \pm i X}$ are the SU(2) generators \Callan .
It is not obvious precisely how we should analytically continue this to
the Lorentzian theory. What is clear is that the analytic continuation
of the $J^\pm$ operators involve the boundary tachyon vertex operators
$e^{\pm t} $.  So after summing over all closed string states
we expect to find that $V_1$ has the form $V_1 =g_s( Ae^t + Be^{-t})$
where $A$ and $B$ are some numerical coefficients. The factor of
$g_s$ comes from the fact that the vertex operator for the emitted
state comes with a factor of $g_s$.
Then the operator $V_B$ will have a form proportional to
$g_s (At e^t -  B t e^{-t})$,
we can interpret this as giving us the
time variation for the coefficient $\lambda$.
In other words, we find $\dot \lambda \sim g_s $.
 Unfortunately we could not find
a more precise equation for the effective evolution of $\hat \lambda$.

Note that if we have a profile such as \rhohalf\ with a $\hat \lambda(t)$
which is a slowly varying function of $t$, this will have the
effect to shift the pole position from ${\rm Im} t_p = \pi $ to
\eqn\polepos{
{\rm Im}(t_p) = \pi \left(1 - {\dot {\hat \lambda}(0)\ov{\hat\lambda}(0)}
\right) + \cdots\ ,
}
where we assumed that derivatives of $\lambda$ are small and we
expanded $\lambda$ to first order in time. So we see that if
$\dot \lambda <0$. Then the position of the pole shifts in such a
way that it makes the integral convergent since the exponential
suppression factor of the form $e^{ - 2 {\rm Im}(t_p) E}$ will be greater
than the factor coming from the density of states. Since
we expect that  $\dot {\hat \lambda}$ is of order $g_s$.
This would give the right
order of magnitude for the cutoff in energy, {\it i.e.} we would get a
total emitted energy of order $1/g_s$.
In other words, the evolution of $\hat \lambda$
 is such that the evolution
of
$\rho(t)$ is slower than the evolution with constant $\hat \lambda$.
That means that the tachyon vertex
 operator would be slightly relevant (as
opposed to marginal).


When we considered the imaginary part of the one loop diagram in
Euclidean space we saw that the divergence came from the open string
stretched between the two closest D-branes. This would suggest that
in order to make the diagram finite we would need to make this distance
slightly bigger than the critical distance. This would suggest that
the Euclidean computation would involve the boundary sine Gordon model
with a potential slightly relevant.

Finally let us note that if we had a decaying unstable lump  in an
 ordinary  weakly coupled field theory it would be enough to solve the
classical equations of motion  to find the state it
decays into at very late times. To the extent that a D-brane is
similar to an ordinary lump solution of string theory one would expect
a similar situation. In this case the precise boundary state is the
classical solution. So one would expect it to encode the final state
that the D-brane decays to. It seems, however,
 that the situation here is not
as straightforward.

\newsec{Conclusion}

In this paper we computed the production of closed string modes produced from
time-dependent rolling tachyon solutions which represent the homegenous decay
(or creation and subsequent decay in the case of the full-brane) of an unstable
D$p$-brane. For $p>2$ the total number of particles and total energy radiated
away was finite. However in these cases a homogenous decay is very unphysical
and D$p$-branes  will decay inhomogenously.
For $p\le 2$ we found that the total energy diverges.
We interpreted this as an indication that
a D0-brane  decays completely into closed string modes, in particular
mostly massive ones. We outlined a few approaches to the back reaction
problem. It would be nice to have a better understanding of it.

\bigskip
\noindent{\bf Acknowledgments}

We would like to thank D.~Gross, M.~Gutperle, G.~Horowitz, N. Itzhaki,
S.~Kachru,
R.~Kallosh, D.~Kutasov, S. Minwalla, G. Moore,
J.~Polchinski, L. Rastelli, N. Seiberg, S.~Shenker, A. Strominger,
W.~Taylor, E. Witten, B.~Zwiebach
for useful discussions. HL wants to thank  M.~Sheikh-Jabbari for help.
HL would also like to acknowledge the Stanford ITP for hospitality
while much of this work was done.

NL and  HL   were supported in
part by DOE grant \#DE-FG02-96ER40949 to Rutgers. JM was supported
in part by DOE grant DE-FG02-90ER40542.

\appendix{A}{Particle production from a source}

In this appendix we give a derivation of various formulae
in the main section based on a second quantized approach.
A closed string field can be written as a linear superposition of
basis states:
\eqn\setupsf{
\ket{\Psi} = \sum_s \ket{\Phi_s} \phi_s \ .
}
Each $\psi_s$, the component of the
vector $\ket{\Psi}$ along the basis vector $\ket{\Phi_s}$,
corresponds to a {\it target space field}.
In the presence of a boundary state the field equation is given by
 \eqn\fulleom{
  (Q + \bar Q) |\Psi \rangle =  |B \rangle
 }
Here $B$ is a vector of ghost number $3$ in the state space of
closed string CFT which
satisfies the conservation law
 \eqn\consvl{
 (Q + \bar Q) \ket{B} = 0 \ .
 }
{}From now on we shall restrict to the cases that the source $\ket{B}$
has the form $\ket{B} = c_0^+ \ket{J}$,
with $\ket{J}$ a state of ghost number two.
Then after fixing the Siegel gauge
$b_0^+ \ket{\Psi} = 0$
we find that
 \eqn\modemo{
 ( L_0 + \bar L_0 )  \ket{\Psi} =  \ket{J} \ .
 }
Expanding $\ket{J}$ in components $J_s$ as in  \setupsf, the equations of
motion \modemo\ takes the familiar field theory form
 \eqn\eumo{
 {1 \ov 2} (\p^2 - m^2_s) \phi_s =  J_s \ .
 }

The closed string fields can be quantized using the standard
method ({\it e.g.} see \zuber).
We assume that  the currents $J_s$ have been switched on only for a
finite interval, we may therefore define the ``in'' and ``out'' field
operators
 \eqn\inout{\eqalign{
 \phi_s(x) & = \phi_s^{in} (x) - {2} \int d^d y \, G_s^{ret} (x-y) \,
 J_s (y) \cr
  & = \phi_s^{out} (x) -  {2} \int d^d y \, G_s^{adv} (x-y) \,
 J_s (y) \ ,\cr
 }}
where $G_s^{ret}$ and $G_s^{adv}$ are standard retarded and
advanced Green functions.
In particular if we expand ( $E_{s,
\vec p} = \sqrt{m_s^2 + \vec p^2}$)
 \eqn\modeexP{\eqalign{
 \phi^{in}_s (x) & =  \int {d^{d-1} p \ov (2 \pi)^{d-1}} {1 \ov
 \sqrt{2 E_{s,\vec p}}}
 \, \left(a_{s,\vec p}^{in} \, e^{i p \cdot x} + (a_{s,\vec
 p}^{in})^{\dagger} \, e^{- i p \cdot x} \right) \ ,\cr
 \phi^{out}_s (x) & =  \int {d^{d-1} p \ov (2 \pi)^{d-1}} {1 \ov
 \sqrt{2 E_{s,\vec p}}}
 \, \left(a_{s,\vec p}^{out} \, e^{i p \cdot x} + (a_{s,\vec
 p}^{out})^{\dagger} \, e^{- i p \cdot x} \right) \ ,\cr
 }}
with
\eqn\acomrels{
[a_{s,\vec p}^{in}, (a_{s,\vec
 p'}^{in})^{\dagger}] = [a_{s,\vec p}^{out}, (a_{s,\vec
 p'}^{out})^{\dagger}] = (2 \pi)^{d-1} \delta^{(d-1)} (\vec p -
 \vec p')
}
then
 \eqn\relato{
 a_{s,\vec p}^{out} = a_{s,\vec p}^{in} - {i \ov \sqrt{2 E_s}}
 g_s \, \tilde J_s (- E_{s,\vec p} , - \vec{p})
 }
where $\tilde J_s(p)$ is the Fourier transform of $J_s(x)$ and
\eqn\Jonshell{
\tilde J_s (E_{s, \vec p}, \; \vec p) = \tilde J_s (p)|_{p^0 = E_{s, \vec
p}}, \qquad
E_{s, \vec p} = \sqrt{m_s^2 + \vec p^2} \ .
}

The S-matrix operator which connects the in- and out-fields and
in- and out-states
\eqn\Smatrix{\eqalign{
 & \phi_s^{out} (x) = \hat S^{-1} \, \phi_s^{in} (x) \, \hat S \cr
 & \ket{\rm out} = \hat S^{-1} \ket{\rm in}, \qquad
    \ket{\rm in} = \hat S \ket{\rm out}
 }}
can be written as
 \eqn\smatr{\eqalign{
 \hat S & = T \exp\left[ - i
  \int d^d x \, \phi^{in} (x) \cdot J (x)
 \right] \cr
 & = : \exp\left[ - i  \int d^d x \, \vec \phi^{in} (x) \cdot \vec J (x)
 \right] : \exp \left[ i \, W \right]
}}
where
 \eqn\atr{
 W = \ha  \biggl \langle B \biggl |
 {b_0^+ c_0^- \ov L_0 + \bar L_0 - i
 \ep} \biggr | B \biggr \rangle
  }
In the above, $T$ and  $::$ denote time and normal ordering
respectively,  $\ket{{\rm in},0}$ is the in-vacuum of the second
quantized theory and
$\vec \phi \cdot \vec J = \sum_s
\phi_s \, J_s$ .

It can be shown from the above that the average number and energy of the
particles emitted by the source are given by
\eqn\defbaN{\eqalign{
 \bar N & = \sum_s \int {d^{d-1} k \ov (2 \pi)^{d-1}} \,
 {1 \ov 2E_s} \, \tilde J_s (- E_s,
  - \vec p) \, \tilde J_s (E_s, \vec p) \cr
 &   = 2 {\rm Im} W =   {\rm Im} \,
 \biggl \langle B \biggl |
 { \, b_0^+  c_0^-  \ov L_0 + \bar L_0 - i
 \ep} \biggr | B \biggr \rangle \cr
 }}
and
\eqn\deftoen{\eqalign{
\bar E & = \ha  \sum_s \int {d^{d-1} p \ov (2 \pi)^{d-1}} \,
 \tilde J_s (- E_s,
  - \vec p) \, \tilde J_s (E_s, \vec p)\cr
 & = {\delta \ov \delta a}  \biggl \langle B \biggl |
\left({\, b_0^+  c_0^- \ov L_0 + \bar L_0}\right)_{ret}  \biggr
| B (a) \biggr \rangle \biggr|_{a=0}
}}
where $\ket{B(a)}$ is obtained from $\ket{B}$ by taking $t \to t+a$,
{\it i.e.}
they are seperated by an interval $a$ in time direction. The subscript ``ret''
indicates that one needs to use the retarded propagator.
Note that in \defbaN\ and \deftoen\ only on-shell physical modes contribute.
This is guarrenteed by the conservation law \consvl, which ensures
the decoupling of gauge modes.

Note that  the probability of going from the in-vacuum to the
out-vacuum ({\it i.e.} the probability of no emission) is
 \eqn\vacpo{
 p_0 = |\bra{0,{\rm out}} {\rm in}, 0\rangle |^2
 = e^{- \bar N} \ .
 }
When $\bar N$ is infinite, the $S$-matrix does not
exist, since every matrix element between in- and out-states
vanishes. In a system with an infinite number of degrees of freedom,
inequivalent representations of the canonical commutation relation
may exist. In certain cases, meaningful physical results can still
be extracted, {\it e.g.} as in the case of ``infrared catastrophy''. In
string theory we also have  to worry  potential
divergences from the Hagedorn density of states.

Equation \smatr\ means that the final state is a coherent state. From
equation \inout\ we have that
\eqn\fielex{
\vev{{\rm in}, 0 | \phi_s (x) |{\rm in}, 0}
= -  {2} \int d^d y \, G_s^{ret} (x-y) \,
 J_s (y)  \ .
}
Similarly
\eqn\fieoex{
\vev{{\rm in}, 0 | \phi_s^{(out)} (x) |{\rm in}, 0}
= -  {2} \int d^d y \, G_s^{-} (x-y) \,
 J_s (y)  \
}
where $G^- = G_{ret} - G_{adv}$. Note \fieoex\ is precisely the classical
radiation field. Finally
\eqn\feyoex{
\vev{{\rm out}, 0 | \phi_s (x) |{\rm in}, 0}
= -  {2} \int d^d y \, G_s^{F} (x-y) \,
 J_s (y)  \
}
where $G^F$ denotes the Feynmann propagator.

\appendix{B}{ Field theory model}

In this appendix we consider Sen's effective field model
\refs{\SenMD,\GarousiTR}
 \eqn\acto{S = -  \int dt \,
 V(T) \sqrt{1 - \dot T^2} = \int dt \, L(t)
 }
for the rolling tachyon. Without loss of generality we will
consider the D0-brane theory and take $V(T) =  M_0 f(T) $
with $M_0$ the D0-brane mass. Note that the tachyon field $T$ used
here is not the same as that used in elsewhere in this paper but differs
by a field redefinition.
We will ``derive'' the tachyon potential
$V(T)$ by matching time-dependent solutions of \acto\ to the half-brane solution
discussed in the main text.

The equation of motion following from \acto\ is
 \eqn\euofm{ \dot T = \sqrt{1 - {V^2 \ov E^2}}\ ,
 }
where $E$ is the conserved energy and the Lagrangian evaluated on the solution is given by
 \eqn\lag{
 L = -  V\sqrt{1 - \dot T^2} = - {V^2 \ov E}\ .
 }
On general grounds
we can  identify the on-shell value of
$- L(t)$ with the partition function on the disk, $M_0 \rho (t)$. For the
half-brane, $E = M_0$ and we find
 \eqn\solh{ \sinh^2 {T \ov 2} = \hat \lam e^t\ ,
 }
and
 \eqn\poten{ f(T) = {1 \ov \cosh {T \ov 2}} \ .}
We note that this potential was also recently used in \refs{\BuchelTJ,\LeblondDB}.

With \poten\ we now would like to find the solution to \euofm\ for
general $E$
 \eqn\eufu{ \dot T = \sqrt{1 - {q^2 \ov \cosh^2 {T \ov
  2}}} \ , \qquad q = {M_0 \ov E} \ .
  }
 When $q > 1$, there is a turning
point $\cosh {T \ov 2} = q $ at which $\dot T =0$. Using time
translational symmetry we can set this point to be at $t=0$. On
the other hand when $ q< 1$, the tachyon can climb to the top of
the potential and we will set this point to be $t=0$.

Equation \eufu\ can be integrated with the above boundary
conditions yielding
 \eqn\fullsol{ \sinh{T \ov 2} = \cases{ a \,
 \cosh{t \ov 2}, & $q > 1$ \cr\cr
                          a \, \sinh{t \ov 2}, & $q < 1$
 } \qquad a = \sqrt{|q^2 -1|}\ .
 }
One thus finds
 \eqn\pot{ - L = {V^2 \ov E} = M_0 \left({1 \ov  1 +
 c  e^t}  + {1 \ov  1 + c e^{-t}} -1\right) = \cases{
 {M_0^2 \ov E} {1 \ov 1 + a^2 \cosh^2 {t \ov 2}} & $q > 1$ \cr\cr
 {M_0^2 \ov E} {1 \ov 1 + a^2 \sinh^2 {t \ov 2}} & $q < 1$
 }\ ,}
with
 \eqn\defpc{ c =  {(q-1)^2 \ov a^2} = \left|{q-1 \ov
 q+1}\right| \leq 1 \ .}
\pot\ is precisely the full-brane profile. Remarkably, despite its
non-linear form,  this effective action admits the general
marginal tachyon deformation, constructed as a linear combination
of a half-brane and time reversed half-brane,  as a solution to
its equation of motion. The $\lambda=\ha$ ($c=1$) case, where the
boundary state vanishes, corresponds in these variables to
$T=\infty$, {\it i.e.} the tachyon is sitting at the closed string
vacuum. Note that while  as a function of $t$ \pot\ reproduces the
string theory result, the relation \defpc\ between $c$ and energy
$E$ is different from that in string theory except for the half
brane limit\foot{That effective field theory of the form \acto\
cannot reproduce the precise form of the stress tensor for the
full-brane derived from
 string theory
was pointed out earlier in~\SenQA\ (see also~\KutasovER). This
appears to suggest that higher derivative terms might be
contributing to the effective action.}.

The potential \poten\ is symmetric under $T \to - T$ and is
bounded below everywhere. Hence it only decribes the well behaved half of the potential
in Bosonic string theory but it also
can be thought of as a model for
non-BPS branes in superstring theory. Our analysis before carries over
trivially to superstring case by a scaling $T \to \sqrt{2} T, \,\,
t \to \sqrt{2} t$, {\it e.g.} we have
 \eqn\posup{ f(T) = {1 \ov \cosh {T \ov \sqrt{2}}}\ .
 }
In other words, to apply equations \solh--\defpc\ to the
superstring case, we should  set $\alpha' = 2$.
When expanding the action \acto\ around $T=0$ using \posup, we
find  $m^2 = - \ha$, precisely that of the open string tachyon on
a non-BPS brane.

\subsec{Static solitons}

We now show the action, when including some
spatial directions, also contain  static solitonic
solutions which have the correct D-brane tension.

Solitonic solutions to \acto\ can be obtained by taking analytic
continuation $ t = - i x$ of the $q > 1$ solution \fullsol,
 \eqn\eucsol{ \sinh {T \ov 2} = a \cos{x \ov 2} \ . }
In the case of D-string, the above solutions correspond to a
periodic array of kinks and anti-kinks sitting at $ x = (2 m +1 )
\pi, m \in \Z$. As $a \to \infty$, the size of the kink becomes
infinitely thin and sharply localized.

The mass of the kink (or anti-kink)  can be obtained by
integration an half period ($2 \pi$) of $x$, we find that
 \eqn\massdd{ M = \int_{0}^{2 \pi}  dx \, V(T) \, \sqrt {1 + T'^2}
 = 2 \pi M_0\ .
 }
Note that this is independent of $q$. Alternatively one can
imagine for \acto\ a single soliton is a configuration with
(see {\it e.g.}~\refs{\LambertHK,\SenTM})
 \eqn\kink{ T  =
 \cases{ -\infty & $x<0$ \cr
             + \infty & $ x>0$
 }}
This again has total energy
 \eqn\massp{
 E = M_0 \int {dT \ov \cosh{T \ov 2}}  =  2 \pi M_0
 }
where we have  approximated $ \sqrt{ 1 +  T'^2} \sim  T'$.
 {} From \massdd\ we see that the deformation from the original D-string
configuration at $T=0$ to the periodic array of kink-anti-kinks
does not cost energy, {\it i.e.} it corresponds to a marginal
deformation. Note the ratio of tensions  matches precisely to that
between D-string and D0-brane given by string theory. In the
superstring theory we have an extra $\sqrt{2}$ in the ratios of
tensions due to the scaling mentioned before.

\subsec{Coupling to closed string modes}

Now we couple the above tachyon system to a set of closed string
free fields propagating in the bulk
 \eqn\totac{ S_{tot} = {1 \ov
 \ep} S_0 [T(t)] +  \int dt \, j(t) \phi(t, x=0) - \ha  \int d^d x
 dt \,\left( (\p \phi)^2 + m^2 \phi^2 \right)\ .
 }
In section two we showed that the couplings of physical closed
string modes to the D-brane are essentially given by the one-point
function of the unit operator in the bulk, {\it i.e.} $\rho (t)$ in
\timpa--\rhohalf. This motivates us to identify $j(t)$ with $ L
[T(t)]$. Note that we have made explicit a small parameter $\ep
\propto g$ in the action and there is no other dependence on $\ep$
in $L [T]$.  Thus the equations of motion become
 \eqn\equmo{ {\p L
 \ov \p T(t)} (1 + \ep K) = {d \ov d t}
 \left({\p L \ov \p \dot T} (1 + \ep  K) \right)\ ,
 }
with
 \eqn\defK{
 K = \int dt' \, G_{ret} (t,0;t',0) \, L (T(t'))\ ,
 }
where $G_{ret}$ is the sum of the retarded propagator for all
closed string modes. Multiplying $\dot T$ to both sides of \equmo\
we find that
 \eqn\mcon{
 {d \ov d t} (H (1 + \ep  K)) = - \ep \,  L {d K \ov d
 t}\ ,
 }
where $H$ correspond to the Hamiltonian of $S_0$. Note that the
energy going into the radiation is given by
 \eqn\raden{
 E_{rad} = \int dt dt' \, L(t)  \, \p_t G_{ret} (t,0;t',0) \, L
 (T(t')) \ .
 }
When integrating \mcon\ over time we get the
modified energy conservation law including the part going into the
radiation.

\appendix{C}{ Brane deformation due to closed string backgrounds }

In this appendix we will discuss a very simple example of the problem
of finding the deformation of a boundary state due to a closed
string deformation of the background.

Let us start with a flat spacetime. Let us consider a
D-brane in the 01 directions. Let us now deform the background
by deforming the metric by
\eqn\metrdef{
ds^2 = ds^2_{flat}  + \epsilon ( dx_1^2 - dx_3^2) e^{ik x_2 +  k x_4}\ ,
}
where $\epsilon \ll 1$.
The deformation grows for large $x_4$, but since the brane is localized
near $x_4$, this is not important for our problem. We view
\metrdef\ simply as the leading term for the expansion of the
metric around $x_4$.

This deformation will translate into the insertion of the vertex
operator
\eqn\vertex{
{\cal O}(z,\bar z) =
- { 1 \over  \pi \alpha'}  \epsilon (\partial X_1 \bar \partial X_1
- \partial X_3 \bar \partial X_3)  e^{ik X_2 +  k X_4}
}
The operator product expansion as $z \to \bar z$ will be of the form
\eqn\opeproex{
{\cal O}(z,\bar z) \sim { 2i \over z - \bar z} V_1 + \cdots\ ,
}
where the ellipsis indicate operators with other conformal weights and
\eqn\vonexp{
V_1 = { 1 \over 2} \alpha'
{ \epsilon} \left( i k { \partial_n X_2 \over \pi \alpha'}  + k
{ \partial_n X_4 \over \pi \alpha'} \right)\ .
}
According to the arguments given in the text the boundary
deformation should obey
\eqn\condbdf{
(L_0- 1)V_B = V_1 ~.}
This motivates us to look for solutions of the form
\eqn\formsol{
V_B = Y^2(X^0,X^1) { \partial_n X_2 \over \pi \alpha'} +
Y^4(X^0,X^1) { \partial_n X_4 \over \pi \alpha'}
}
where \condbdf\ implies
\eqn\boundfgh{
(-\partial_0^2 + \partial_1^2 ) Y^2 = {1\over 2} \epsilon ik  ~,~~~~~~~~~
(-\partial_0^2 + \partial_1^2 ) Y^4 = {1\over 2} \epsilon  k\ .
}
Note that $V_B$ and $V_1$ form a ``logarithmic''
pair~\refs{\GurarieXQ,\FlohrZS,\GaberdielTR}.
(The fact that we get a complex answer in \boundfgh\ is due
to the fact that we took a complex metric in \metrdef , taking a real
metric we get  a real answer.)

Let us now compare with the results we would compute using the Born
Infeld action. The action can be expanded to lowest order
in small fluctuations $Y^2,Y^4$ about $X^m=0$ as
\eqn\action{
S = -T\int dx^0dx^1 \sqrt{-g}
=- T \int dx^0 dx^1\left( 1+ { 1 \over 2}( \partial Y^2)^2    + { 1 \over 2}( \partial Y^4)^2    + { 1 \over 2}
\delta g_{11}\right)\ .
}
The equations of motion for this action become \boundfgh\ once we
use \metrdef\ for the expression for $\delta g_{11}=\epsilon e^{ikY^2+kY^4}$.
We see that indeed the general CFT computation agrees with the answer
we expect based on the Born Infeld action. Note that since the
\metrdef\ is independent of $x^0, x^1$ the solutions of \boundfgh\ will
grow at infinity. This is in agreement with the expectation that
logarithmic divergences in CFT are cancelled by a change in the
long distance boundary conditions for the configuration.

It should be noted that if we give the metric deformation some
momentum along the brane. Then the operators in the OPE will not
generally have dimension one. It is not clear if we need to deform the
boundary state in this case since we do not get a logarithmic divergence.
 {}From the Born Infeld point of view we
would expect that we need to change the boundary state.

This discussion is closely related to the discussions
of brane recoil in \refs{\FischlerJA,\PO,\KMW}.

\listrefs

\bye

\listtoc
\writetoc

\bye